\documentclass[a4paper,10pt]{article}
\usepackage[utf8]{inputenc}
\usepackage{amsmath}
\usepackage{amssymb}
\usepackage{latexsym}
\usepackage{color}
\providecommand{\Compile}[1]{}
\newcommand{\cosec}{\mbox{\rm cosec}}
\title
{\bf Shape Invariant Rational Extensions And Potentials Related to Exceptional
Polynomials}

\author{
S. Sree Ranjani$^{a,b}$, R. Sandhya$^{c}$ and A. K. Kapoor$^{d,e}$\\
$^a$School of Physics, University of Hyderabad,\\ Hyderabad, India, 500
046.\\[3mm]
$^{b*}$Faculty of Science and Technology,\\ ICFAI foundation for Higher
Education,\\
(Declared as Deemed-to-be University u/s 3 of the UGC Act 1956)
\\ Dontanapally, Hyderabad, India, 501203 \footnote{ Present
Address}$^*$.\\[3mm]
$^c$Department of Physics, Osmania University, Hyderabad, India, \\[3mm]
$^d$Department of Physics, Shiv Nadar University,\\
Village Chithera, Tehsil Dadri, Gautam Budh Nagar,\\
UP, India, 203207
.\\[3mm]
$^{e*}$ Department of Physics, BITS-Pilani Hyderabad Campus\\
Jawahar Nagar, Shameerpet Mandal,\\
Hyderabad 500078, India
}
\begin{document}
\maketitle

\begin{abstract}
In this paper, we show that an attempt to construct shape invariant extensions of a known shape invariant potential leads to, apart from a shift by a constant, the well known technique of isospectral shift deformation.  Using this, we construct infinite sets of generalized potentials with $X_m$ exceptional polynomials as solutions.
These potentials are rational extensions of the existing shape invariant
potentials.  The method is elucidated using the radial oscillator and the trigonometric
P\"{o}schl-Teller potentials. For the case of radial oscillator, in addition to the known rational extensions, we construct
two infinite  sets of rational extensions, which seem to be less studied. For one of the potential, we show that its solutions involve a third type of exceptional Laguerre polynomials. Explicit expressions of this generalized infinite set of potentials and the corresponding solutions are presented. For the trigonometric P\"{o}schl-Teller potential, our analysis points to the possibility of several rational extensions beyond those known in literature.\\
\end{abstract}

\newpage

\section{Introduction}

     The search for new exactly solvable (ES) models gained momentum with the
discovery of the exceptional orthogonal polynomials (EOPs)
\cite{kam1},\cite{kam2}. New potentials with the $X_1$ EOPs as solutions were
constructed in \cite{quesne},\cite{sasaki}. They turned out to be the rational
extensions of the well studied one dimensional bound state problems namely the radial oscillator and
the Darboux P\"oschl-Teller (DPT) potentials \cite{khare_book}. A spate of
papers followed, where generalized rational extensions of many one dimensional
shape invariant potentials (SIPs) were presented \cite{kam3}- \cite{chait}.
Different methods like the Darboux-Crum method \cite{sas_crum}, \cite{kam_dar}, 
finite difference Backl\"und algorithm \cite{gran1}-\cite{gran_dbt}, prepotential 
method \cite{ho} and several others \cite{dutta}, \cite{chait} were employed. 
Various groups explored the mathematical properties of the new polynomials concurrently
\cite{sasaki_sigma} - \cite{gomez2}.

     Many studies showed that the traditional SIPs and their rational extensions
are isospectral and share a supersymmetric partnership \cite{quesne},
\cite{dutta}, \cite{chait}.  In recent papers, Grandati {\it et.al.,}
\cite{gran1}, \cite{gran2} constructed rational extensions of SIPs using
regularized  excited states Riccati-Schr\"odinger functions as superpotentials.
The finite difference B\"acklund transformation was combined with the
regularization scheme to generate  infinite
sets of rational extensions of the isotonic oscillator \cite{gran2}.
Dutta, Roy and coworkers \cite{dutta}, \cite{ces} and references therein,
constructed new conditionally exactly solvable (CES) models using supersymmetric
techniques. Among these a particular construction gave potentials related to exceptional
polynomials \cite{ces}.

Shape invariance (SI) property, discovered in \cite{ged}, has been central to the
discussion of the exact solutions of potential problems in supersymmetric
quantum mechanics (SUSYQM) \cite{khare_book}, \cite{wit}, \cite{coop}.
 We make this property the focus of our study in
this paper. Here we formulate a sufficient condition for SI  using the quantum Hamilton Jacobi (QHJ) formalism \cite{lea} - \cite{qes}.
This condition brings out the fact that SI is a property of a potential 
and does not require knowledge or a reference to a SUSY partner potential or the 
superpotential. This requirement makes it possible to analyze the SI 
property and use it to obtain deformed potentials which are shape invariant by construction. This
will be illustrated using the radial oscillator and the Darboux P\"oschl-Teller (DPT) potentials.
In an earlier paper \cite{sip}, it has been shown that all previously known SIPs, as listed in 
\cite {khare_book}, can be derived in an extremely simple,
straightforward method using the sufficient condition for SI. 
In this paper we complete this program by constructing rational SIPs of radial oscillator
leading to potentials related to the $L1$ and $L2$ exceptional polynomials \cite{sigma}, \cite{sas_crum}, \cite{kam_dar}, \cite{gran2}, \cite{gran_dbt}. In addition, we also find one more 
 generalized set of potentials with polynomial solutions, which may be classified as
type III or $L3$ exceptional polynomials \cite{sigma}, \cite{gran2}. The expressions for these generalized potentials and their polynomial solutions are presented here for the first time. For the DPT potential, we are led to many rational extensions of which some coincide with the known extensions in the literature \cite{quesne}, \cite{sigma}, \cite{sas_crum}.  

Our scheme allows us to construct a hierarchy
of generalized  ES rational potentials indexed by the hierarchy index $m$,
$(m=0,1,2\dots)$. Each value of $m$ gives a new rational extension of the
original SIP and the solutions are in terms of the
$X_m$ EOPs. For $m=0$, the generalized potentials reduce to the original SIPs.
In the scheme presented in this paper, the superpotentials of the original SIP,
corresponding to both exact and broken SUSY,  play a central
role. The interplay of ideas and the use of tools available in the QHJ formalism
and SUSYQM makes our scheme very simple, elegant  and transparent.

The knowledge of the singularity structure of the quantum momentum function
(QMF) within the QHJ formalism, allows one to construct all possible
superpotentials associated with a given supersymmetric potential. Next for each
superpotential, say $W(x)$,  we find extended superpotential, $\widetilde{W}(x)$, such that the corresponding potentials $\widetilde{V}^{(+)}(x)$ is shifted w.r.t. $V^{(+)}(x)$  by a constant. Each of the resulting  potentials, $\widetilde{V}^{(-)}(x)$, associated with $\widetilde{W}(x)$, is either a singular or a regular rational extension of $V^{(-)}(x)$. This method has been earlier used in \cite{dutta} to obtain CES rational potentials with EOPs as solutions. The superpotentials associated with nonnormalizable
solutions of $V^{(-)}(x)$ lead to the regular extensions and these potentials have eigenfunctions in terms of the $X_{m}$ exceptional polynomials as solutions. The potentials associated with the normalizable wave functions result in new sets of generalized potentials which are singular or regular depending on the potential parameters. Some of these potentials are discussed in \cite{sigma}, \cite{gran2}.

In Sec. 2 we first recall some important features of the QHJ formalism of quantum mechanics and reformulate the SI requirement, crucial to our discussion.  In Sec.3, we show that this requirement leads to the isospectral shift deformation of the SIPs known from the early days
of SUSYQM. Sec.4 has a detailed discussion of rational extensions of the radial
oscillator. We also present the expressions for an infinite set of rational potentials, whose solutions are in terms of the new set of polynomials, the $L_3$ category of exceptional Laguerre polynomials. The case of trigonometric DPT, leading to the exceptional Jacobi polynomials is taken up in Sec.5. Finally we present our conclusions in Sec.6.

\section{The QHJ and SUSYQM formalisms}
In the QHJ formalism \cite{lea} - \cite{the}, the starting point is the Riccati equation 
\begin{eqnarray}
     p^2(x,E) - i\hbar \frac{dp(x,E)}{dx} = 2m (E-V(x)),  \label{E1.1}
\end{eqnarray}
obtained by making the substitution $\psi(x,E)=\exp(iS(x,E)/\hbar)$ in the 
Schrödinger equation. Here $p(x,E)$, the quantum momentum function
(QMF), is identified with $\frac{dS(x,E)}{dx}$. For convenience, we will use units so that $\hbar=1,
2m=1$ and in order to  work with real equations, we define $Q(x,E)$ by  $p(x,E) \equiv i Q(x,E)$. 
The QHJ equation for $Q(x,E)$ assumes the form 
\begin{equation}
    Q^2(x,E) - \frac{dQ(x,E)}{dx} = V(x) - E .\label{E1.2}
\end{equation}
$Q(x,E)$ is simply the negative of the logarithmic derivative of solution 
$\psi(x,E)$ of the Schr\"{o}dinger equation
\begin{equation}
   Q(x,E) =- \frac{d}{dx} \log \psi(x,E) =-
\frac{1}{\psi(x,E)}\frac{d\psi(x,E)}{dx}
\label{E1.3}.
\end{equation}
The QHJ approach essentially makes use of the  singularities of the function
$Q(x,E)$ in the complex $x$ plane. A crucial step in solving for the QMF, $Q(x,E)$, is to
write it as a sum of two parts,
\begin{equation}
   Q(x,E) = Q_\text{f} + Q_\text{m} \label{E1.4}
\end{equation}
where $Q_\text{f}$ contains the fixed singularities and its form can be written
down by inspection of different terms in the QHJ equation. The second part,
$Q_\text{m}$, contains the singular part having the moving singularities which depend on the 
boundary conditions to be imposed. In all known cases of ES potentials \cite{es}, \cite{the} including the new rational potentials with EOP solutions \cite{sree_eop} and the quasi-exactly solvable potentials \cite{qes}, when $E$ is taken to be one of the energy
eigenvalues, $Q_m$ turns out to have a {\it finite number of simple
poles with residue -1}. It is easy to see from (\ref{E1.3}) that  the moving poles of
$Q(x)$ are located at the nodes and at other complex zeros of the wave
function.

The superpotential $w(x)$, which  in SUSYQM corresponds 
to  the negative of logarithmic derivative of a solution of the 
Schrödinger equation, plays an important role of defining the partner potential and 
formulating the SI condition \cite{ged}.  Usually, the solution chosen
has no nodes but may, or may not, be normalizable, the two cases being referred to as cases 
of exact and broken SUSY respectively. Thus $w(x)$ happens to be just one of the
solutions $Q(x, E)$ of the QHJ equation, for certain energy, for which moving poles are
absent and $Q_\text{m}$  in (\ref{E1.4}) is replaced by a constant. The SUSY partner 
potentials $V^{\pm}(x)$ are defined by 
\begin{equation}
 V^{(\pm)}(x) = w(x)^2 \pm w^\prime(x)\label{E1.5}
\end{equation}
and a potential $V(x)$ is defined to be shape invariant if  both the partner
potentials have the same form as $V(x)$ apart from an overall constant. Since 
several superpotentials are possible, hence the
definition of partners, and the property of SI itself appears to
depend on the choice of the superpotential. It turns out to be extremely useful to
formulate SI as a property which does not require the use of the
superpotential, and which makes it  transparent as a property of
the potential.  As we will see later, this fact assumes
importance in the context of rational extensions related to the exceptional
polynomials.

We now reformulate the SI requirement on a potential
$V(x)$. Let   $Q(x,\sigma)$ be a solution of the QHJ equation,
\begin{equation}
    Q^2(x,\sigma) - \frac{dQ(x,\sigma)}{dx} = V(x) - E(\sigma)
\label{E1.9},
\end{equation}
where the dependence of $Q(x,\sigma)$ on the potential parameters has been
explicitly shown through constants $\sigma$, which  are functions of the
parameters appearing in the potential. If we introduce
two functions $w_1(x), w_2(x)$
\begin{eqnarray}
   w_1(x) = Q(x, \lambda), \qquad  w_2(x) = Q(x,\mu).\label{E1.6A}
\end{eqnarray}
corresponding to two different values $\sigma=\lambda, \mu$, then the two
potentials, defined by
\begin{equation}
 V_1(x) = w_1^2(x)-w_1^\prime(x), \qquad V_2(x) = w_2^2(x) - w_2^\prime(x),
 \label{E1.7}
\end{equation}
obviously will have the same shape. If we further require that there
exists a map
\begin{equation}
  \tau : \lambda \to \mu=\tau(\lambda)
\end{equation}
such that
\begin{equation}
 Q(x,\lambda) + Q(x,\mu)=0,  \label{E1.8}
\end{equation}
then $w_2(x) =- w_1(x)$ and the two potentials $V_{1,2}(x)$ will be SUSY
partners.

{\it
To summarize, given a potential $V(x)$, the twin requirements of existence of
solutions
$Q(x,\lambda)$, $Q(x,\mu)$ of  QHJE and of a map $\tau$ such that
\begin{equation}
    Q(x, \mu) = - Q(x, \tau(\lambda))\label{Eq1.9}
\end{equation}
are seen to be the  {\it sufficient} conditions for SI of the
potential $V(x)$. }


\section{Shape Invariant Rational Extensions}
Given a superpotential $w(x,\lambda)$, the SI requirement means the
potential $V^{(+)}(x,\lambda)= w^2(x,\lambda)+w^\prime(x,\lambda)$ be equal to $V^{(-)}(x,
\mu)\equiv w^2(x,\mu)-w^\prime(x,\mu)$, apart from an additive constant, after the set of potential parameters,
$\lambda$, is redefined. This translates into the following requirement on the superpotential as
\begin{equation}
     w^2(x,\lambda) + w^\prime(x,\lambda) =   w^2(x,\mu) - w^\prime(x,\mu)
         +\text{constant}.\label{E2.1}
\end{equation}
for some $\lambda, \mu$ and a constant which may depend on $\lambda,\mu$.

Let us now assume that a particular superpotential $w_0(x,\lambda)$ has been
found which gives rise to a SIP $V(x,\lambda)$, where $\lambda$ denotes  constants
which are functions of the potential parameters.
We  now seek to construct a shape invariant extension of $w_0(x, \lambda)$ by constructing an extended superpotential
\begin{equation}
    \widetilde{w}(x, \lambda) =  w_0(x, \lambda) + \phi(x,\lambda)
\label{E1.6B},
\end{equation}
where $\phi(x,\lambda)$ is an unknown function and is to be determined from 
the requirements of SI.

Using the fact that both $ \widetilde{w}(x)$ and $w_0(x)$ for some $\mu, \lambda$,
satisfy the SI requirement (\ref{E2.1}), we get
\begin{eqnarray}
\lefteqn{
   \phi(x,\lambda)^2 + 2 w_0(x,\lambda) \phi(x, \lambda) + 
\phi^\prime(x,\lambda)} &&  \\ 
&=&   \phi(x,\mu)^2 + 2 w_0(x,\mu) \phi(x, \mu) - \phi^\prime(x,\mu) + K, \label{E2.2}
\end{eqnarray}
where $K$ is a constant. For SIPs of interest in this paper, and defined by
$w_0(x,\lambda)$, there exists a map
$\tau : \lambda \to \mu = \tau(\lambda)$ such that
\begin{equation}
    w_0(x,\mu) = - w_0(x, \tau(\lambda)).\label{EQ14}
\end{equation}
We also want the extended potential to be shape invariant and hence we demand that
\begin{equation}
\phi(x,\lambda) \rightarrow -\phi(x, \tau(\lambda)).
\end{equation}
Thus we can rewrite (\ref{E2.2}) as
\begin{eqnarray}
\lefteqn{
    \phi(x,\lambda)^2 + 2 w_0(x,\lambda) \phi(x, \lambda) + 
\phi^\prime(x,\lambda)}   && \\ 
  &=&   \phi(x,\tau(\lambda))^2 + 2 w_0(x,\tau(\lambda)) \phi(x, \tau(\lambda))
+ \phi^\prime(x,\tau(\lambda)) + K.
\label{EQ16}
\end{eqnarray}
The structure of the above equation suggests using the ansatz of equating both 
sides of the above equation to a constant. Therefore, we look for solutions for 
$\phi(x,\lambda)$ satisfying the equation
\begin{equation}
    \phi(x,\lambda)^2 + 2 w_0(x,\lambda) \phi(x, \lambda) + 
        \phi^\prime(x,\lambda) =   R_1,  \label{EQ17}
\end{equation}
where $R_1$ is a constant depending on the potential parameters $\lambda$.
The above condition, (\ref{EQ17}), is simply the relation
\begin{equation}
      \widetilde{V}^{(+)}(x) = V^{(+)}(x)+ R_1, \label{EQ18}
\end{equation}
where the notation $\widetilde{V}^{(\pm)}(x)$ and $V^{(\pm)}(x)$ denote the SUSY
partner potentials constructed from $\widetilde{w}(x)$ and $w_0(x)$
respectively. It may be remarked that we have recovered the property that has
been used in \cite{dutta} to construct new conditionally exactly
solvable models with solutions related to the exceptional polynomials.
For the special choice $R_1=0$, this requirement is same as that used in 
construction of strictly isospectral deformations of a given potential \cite{khare_iso}, \cite{sukatme_iso},
\begin{equation}
    V^{(-)}(x) = w_0^2(x, \lambda) - w_0^\prime(x, \lambda).\label{EQ18A}
\end{equation}
In general the constant $R_1$ in \eqref{EQ18} can be non zero and  the above
process of defining $\widetilde{V}^{(-)}(x)$ by means of
\begin{equation}
      \widetilde{V}^{(-)}(x) = \widetilde{w}^2(x,\lambda) -
\widetilde{w}^\prime(x,\lambda) \label{EQ183}
\end{equation}
will be called the {\it isospectral shift deformation} of potential $V^{(-)}(x)$.

The equation for $\phi(x,\lambda)$ is the Riccati equation and can be linearised by introducing
$u(x,\lambda)$ defined by
\begin{equation}
 \phi(x, \lambda)=\frac{1}{u(x,\lambda)}\frac{d u(x,\lambda)}{dx} \label{EQ18B}.
\end{equation}
The unknown function
$u(x,\lambda)$ then satisfies a Schr\"{o}dinger like equation
\begin{equation}
 \frac{d^2 u(x,\lambda)}{dx^2} + 2 w_0(x,\lambda) u(x,\lambda) - R_1
u(x,\lambda) =0. \label{EQ19}
\end{equation}
In order to obtain rational extensions and to connect with the exceptional
polynomials, the constant $R_1$ will be chosen so as to give
polynomial solutions for $u(x)$. Let $P_m(x)$ denote a polynomial solution, of
degree $m$, of \eqref{EQ19}. A  rational  extension of the original SIP $V(x)$, 
will be obtained from the   polynomial solutions $P_m(x)$ for $u(x)$ by means of 
the relations
\begin{equation}
  \widetilde{V}^{(-)} (x) = \widetilde{w}^2(x) - \widetilde{w}^{\,\prime}(x),
\label{EQ20}
\end{equation}
where
\begin{equation}
       \widetilde{w}(x) = w_0(x) + \phi(x),\qquad \phi(x) = \label{EQ21}
\frac{1}{P_m(x)}\frac{d P_m(x)}{dx}.
\end{equation}
The equations (\ref{EQ19}) - (\ref{EQ21}) are the basic equations for the
rational extensions  and to arrive at the Hamiltonians which have eigenfunctions
in terms  of the EOPs.

In general when $R_1$ is not restricted and is allowed real values, we would
get an extended potential which would interpolate between the potentials
related to the exceptional polynomials.

\section{Rational Extensions of the Radial Oscillator}
\subsection{Solutions for the Superpotential}
In this section we will construct the rational extensions of the radial
oscillator given by ( $2m=1$)
\begin{equation}
  V(r) = \frac{1}{4}\omega^2 r^2 + \frac{\ell(\ell+1)}{r^2}.
\label{EQ4.1}
\end{equation}
Four superpotentials $w(r) $ for the radial oscillator,  obtained from the
solutions of the QHJ equation,
\begin{equation}
     w^2(r)- w^\prime(r) = V(r) -E \label{Eq4.2},
\end{equation}
and the corresponding partner potentials $V^{(\pm)}(r) $ are given in the table
below.\\
\begin{center}
      {Table 1 }
\end{center}
\begin{equation}
\begin{array}{|c|l|l|l|}\hline &&&\\[0.5mm]
k  & \text{Superpotential}~ w_k
    & V^{(-)}_k= w^2_k-w^\prime_k
    &  V^{(+)}_k= w^2_k + w^\prime_k\\[3mm]    \hline &&&\\
 1  & \dfrac{1}{2}\omega r-\dfrac{(\ell+1)}{r}
    & V(r)- \omega(\ell+3/2)
    &\dfrac{1}{4}\omega^2 r^2 +\dfrac{(\ell+1)(\ell+2)}{r^2}- \omega(\ell
            +{1}/{2})  \\[3mm]\hline &&&\\
 2  &\dfrac{1}{2}\omega r  + \dfrac{\ell}{r}
    & V(r)+ \omega(\ell-1/2)
    & \dfrac{1}{4}\omega^2 r^2 + \dfrac{\ell(\ell-1)}{r^2} +
\omega(\ell+1/2)\\[3mm] \hline &&&\\
 3  & -\dfrac{1}{2}\omega r -\dfrac{(\ell+1)}{r}
    &  V(r)+\omega(\ell+3/2)
    &  \dfrac{1}{4}\omega^2 r^2 + \dfrac{(\ell+1)(\ell+2)}{r^2}+\omega(\ell
            +{1}/{2})  \\[3mm]\hline &&&\\
 4  & -\dfrac{1}{2}\omega r  + \dfrac{\ell}{r}
    &  V(r)-\omega(\ell-1/2)
    & \dfrac{1}{4}\omega^2 r^2 + \dfrac{\ell(\ell-1)}{r^2}-\omega(\ell
            +{1}/{2})\\[3mm]  \hline
\end{array}  \label{Eq4.6}
\end{equation}

These four solutions of the QHJ equation are meromorphic, have no moving poles
and are
of the form
\begin{equation}
  w(r,\sigma) = \frac{a}{r} + \frac{1}{2} b r,\qquad \sigma=\{a,b\},
  \label{Eq4.7}
\end{equation}
where $a,b$ have two possible values
\begin{equation}
      a= -(\ell+1), \,\ell; \qquad b= \pm \,\omega. \label{Eq4.8}
\end{equation}
The four superpotentials correspond to the four possible combinations of
values for $\{a,b\}$. The parameter values,
\begin{equation}
    \mu=\{-\ell-1, \omega \}, \quad       \lambda=\{\ell, -\omega\}.
\label{Eq4.10}
\end{equation}
correspond to the two solutions $<1>$ and $<4>$ in \eqref{Eq4.6}
of QHJE.

With translation mapping $\tau$ defined as
           $$\tau (\{\ell,\omega \}) = \{\ell+1,\omega\} $$
we have $\tau(\lambda) = \{\ell+1, -\omega\}$.
The SI property (\ref{Eq1.9}),
\begin{equation}
    w(r, \mu) = -w(r,\tau(\lambda)), \label{Eq4.9}
\end{equation}
for the radial oscillator, is obviously satisfied.
The solutions $w_2(x), w_3(x)$ for the superpotential, in the second and the
third rows of the table correspond to broken supersymmetry. These also are
related by a relation of the form \eqref{Eq4.9}.  These will be used for
constructing rational potentials related to the $L_1$ and $L_2$ type
exceptional Laguerre polynomials.

The extended potentials corresponding to $w_1(r)$ and $w_4(r)$ are also
constructed. Most of these potentials turn out to be singular. However in some
cases regular rational extensions are found and the corresponding polynomial
solutions are presented in Sec 4.4.
\subsection{Isospectral Shift Deformation}
The isospectral shift for the radial oscillator will be used to construct a new
potential $\widetilde{V}(r)$ by making use of Eqs.
(\ref{EQ18A})-(\ref{EQ21}). Identifying  $w_0(r)$ with $w_2(r)$, we introduce
\begin{equation}
  \widetilde{w}_2(r)= w_2(r) + \phi(r),
\end{equation}
and write $\phi(r) = \frac{d}{dr}\log u(r)$, the equation
(\ref{EQ19}) then becomes
\begin{equation}
   u{''}(r) + \Big( \omega r + \frac{2\ell}{r}\Big) u{'}(r) - R u(r)
=0.\label{Eq36}
\end{equation}
That this equation reduces to the Laguerre  equation, is most easily seen by
making a change of  variable $\eta = -\frac{1}{2}\omega r^2$ which transforms
the above equation to
\begin{equation}
  \eta u{''}(\eta) + \Big(-\eta +\ell+\frac{1}{2}\Big) u{'}(\eta) +
\frac{R}{2\omega} u(\eta) =0. \label{EQ37}
\end{equation}
Demanding that the solution for  $u(\eta)$ be a polynomial in $\eta$ of degree
$m$, gives $R=2m\omega$ and the polynomial solution coincides with the
associated Laguerre polynomials \cite{den}, \cite{ince} and we have
\begin{equation}
   u(\eta) = L_m^{\alpha}(\eta),  \qquad \phi(r) = -\omega r
   \frac{\partial_\eta L_m^{\alpha}\eta)}{L_m^{\alpha}(\eta)},
   \qquad\alpha=(\ell-1/2).
\end{equation}
Therefore, indicating the dependence of the solutions on $m$, the hierarchy
index explicitly, we write
\begin{equation}
\widetilde{w}_{2,m}(y) = w_2(y) + \omega r \frac{\partial_y
L_m^{\alpha}(y)}{L_m^{\alpha}(y)}.   \label{a}
\end{equation}
where  $y=\frac{1}{2}\omega r^2=-\eta$ is the variable that will appears  in
all final expressions.  Substituting \eqref{a} in \eqref{EQ20}
gives the final expression for the rational extension as
\begin{equation}
  \widetilde{V}_m^{(-)}(r) = V^{(-)}(r)  +2\omega r\left[\omega r
+\frac{2\ell}{r}\right]
\frac{\partial_yL^{\alpha}_m(y)}{L^{\alpha}_m(y)
} -\omega r\partial_y\left[\omega
r\frac{\partial_yL^{\alpha}_m(y)}{L^{\alpha}_m(y)}\right] +
\left(\omega r\frac{\partial_yL^{\alpha}_m(y)}{L^{\alpha}
_m(y) }\right)^2.   \label{vtil_RO}
\end{equation}
$\widetilde{V}_m^{(-)}(r)$ to show the dependency on the hiararchy index $m$.
For $m=0$ all the rational expressions reduce to the undeformed potentials. For
each value of $m>0$, we get a distinct rational extension of the radial
oscillator. The partner potential $\widetilde{V}^{(+)}(r)$
\begin{equation}
  \widetilde{V}^{(+)}(r)= \frac{1}{4}\omega^2 r^2 +\frac{\ell(\ell-1)}{r^2}+2 m
\omega,
\label{vplustil_RO}
\end{equation}
 by construction, coincides with the radial oscillator. The eigenfunctions of
$\widetilde{V}^{(-)}_m(r)$
can be obtained by using the (broken) SUSY intertwining relation  between the
solutions of the partners, which for this case becomes
\begin{eqnarray}
\widetilde{\psi}_{n,\,m}^{(-)}(y)&=&\left( -\frac{d}{dr}
+\widetilde{w}_{2m}(r)\right)\widetilde{\psi}^{(+)}_n(r), \label{EQ341}
\end{eqnarray}
where 
\begin{eqnarray}
\widetilde{\psi}_{n}^{(+)}(r)
&=&y^{\ell/2}\exp(-y/2){L}_n^{\alpha}(y)|_{y=\frac{1}{2}\omega r^2},
\label{EQ342}
\end{eqnarray}
are obtained by replacing $\ell\to \ell-1$ in the eigenfunctions of
the radial oscillator \eqref{A03} in the appendix. On simplification, as
explained in the appendix, \eqref{EQ341} gives
\begin{eqnarray}
\widetilde{\psi}_{n,\,m}^{(-)}(r)=\left[\frac{r^{\ell/2}\exp(-\frac{1}{4}\omega
r^2)}{L^\alpha_m(-y)
} \left({L}_m^{(\alpha+1)}(-y)
{L}_n^{\alpha}(y)  - {L_m^{\alpha}(-y)}{\partial_y
L_n^{\alpha}(y)}\right)\right]_{y=\frac{1}{2}\omega r^2}.
  \label{psithil}
\end{eqnarray}
The recurrence relation\cite{den}
\begin{equation}
\frac{d}{dr}L^{\alpha}_n(r)= L^{\alpha}_n(r)- L^{\alpha+1}_n(r)
\label{recur}
\end{equation}
has been used to write the eigenfunction in the form \eqref{psithil}.
The details of this step can be found in the appendix. The eigenfunctions of
the extended potential have the form
\begin{equation}
\widetilde{\psi}^{(-)}_{n+1,\,m}(r)=\left(\frac{r^{\ell/2}\exp\big(-\frac{1}{4}
\omega r^2\big)}{L_m^{\alpha}(-y)|_{y=\frac{1}{2}\omega
r^2}}\right)\widetilde{P}_{n,m}(r),              \label{RO_ext_sol}
\end{equation}
where
\begin{equation}
\widetilde{P}_{n,m}(r)=\left[{L}_m^{(\ell+\frac{1}{2})}(-y)
L_n^{(\ell-\frac{1}{2})}(y)-L_m^{(\ell-\frac{1}{2})}(-y){\partial_y
L_n^{(\ell-\frac{1}{2})}(y)}\right]_{y=\frac{1}{2}\omega r^2}.\label{L1Exp}
\end{equation}
The polynomials $\widetilde{P}_{n,m}(r)$ are the $X_m$ EOPs, orthogonal with
respect to the weight function
\begin{equation}
\mathcal{W}(r)=\exp\left(-\int \widetilde{w}_{2,m}(r) dr\right) =
\frac{r^{\ell/2}\exp\left(-\frac{1}{4}\omega
r^2\right)}{L_m^{\alpha}(-\frac{1}{2}\omega r^2)},  \label{new_wtf}
\end{equation}
in the interval $0<r<\infty$. The above polynomials have $n+m$ zeros, where $n$
real zeros are inside the interval of
orthogonality and $m$ zeros, either real
or complex, lie outside the interval. For all the states, the number of zeros
outside the orthogonality interval remains fixed, while the number of real
zeros located in the interval is governed by the oscillation theorem. These
polynomials
in \eqref{L1Exp} reduce to 'undeformed polynomials' $L_n^{\ell+1/2}(\omega
r^2/2)$ when $m$ is set equal to zero.
From the above, one can see that each value of $m$ gives a different rational
extension of the radial oscillator \eqref{EQ4.1}, whose $n^\text{th}$ excited
state
wave function can be obtained from \eqref{RO_ext_sol}.   For example, putting
$m=1$ in the equations \eqref{vtil_RO} and simplifying gives
\begin{equation}
\widetilde{V}^{(-)}_1(r)= \frac{1}{4} \omega^2 r^2 +\frac{\ell(\ell+1)}{r^2} +
\frac{2\omega^2 r^2+4\omega \ell -2\omega }{\omega r^2 +2\ell+1}  +\frac{8
\omega^2
r^2}{(\omega r^2 +2\ell+1)^2} + (\ell-\frac{1}{2})\omega
\end{equation}
which matches with the potential constructed by Quesne in \cite{quesne}.
solutions.
Similarly substituting $m=1$ in \eqref{psithil}, we obtain
\begin{equation}
\widetilde{\psi}_{n,\,1}^{(-)}(r)=\left(\frac{r^{\ell+1}\exp\left(-\frac{1}{4}
\omega
r^2\right)}{ \omega r^2+2\ell+1}\right)\left[{(\omega
r^2+2\ell+1)}{L}_n^{(\ell+\frac{1}{2})}(\frac{1}{2}\omega r^2) + 
{L}_n^{(\ell-\frac{1}{2})}(\frac{1}{2}\omega r^2)\right],
\end{equation}
where the polynomial part coincides with the $X_1$ exceptional Laguerre
polynomials of $L_1$ series \cite{quesne}, \cite{sas_crum}.  \\

\subsection{Shape invariance of the extended potential}
We arrived at the new potential $\widetilde{V}^{(-)}(r)$ by analysing the SI
condition \eqref{E2.1} for the extended potential. However,
at this stage the SI of the extended potential, though guaranteed, is not
transparent. This  is  because the SUSY partner  of
$\widetilde{V}^{(-)}(r)$ obtained using $\widetilde{w}_{2,m}$ is, by
definition, same as potential $V^{(+)}(r)$ apart from a constant.
In order to explicitly demonstrate the SI, we obtain a new
superpotential satisfying the twin properties given at the end of Sec, 2.
For this purpose, we now start with $w_3(r)$ and carry out a similar process
with roles of partners exchanged. Specifically, we construct a rational
extension $\overline{w}(r)=
w_3(r) + \chi(r)$ and carry out isospectral shift deformation, exchanging
roles of $V^{(\pm)}(r)$. In this case, the solution  for $\chi(r)$ is obtained
by demanding
\begin{equation}
  \overline{V}^{(-)}(r) = V^{(-)}(r)  + R_2 \label{EQ501}
\end{equation}
and construct an extended potential
\begin{equation}
    \overline{V}^{(+)}(r) = \overline{w}(r)^2 + \overline{w}^\prime(r),
\label{EQ502}
\end{equation}
which turns out to be the SUSY partner of $\widetilde{V}(r)$ obtained
in the previous section. Going through the same steps as above, the equation for
$\bar{w}(r)$ turns
out to be
\begin{equation}
    \chi^2(r) + 2w_3 (r) \chi(r) -\chi^\prime(r) =R_2,
\end{equation}
where  $R_2$ is a constant depending on the potential parameters $\lambda$ and 
\begin{equation}
  \overline{w}(r) = w_3(r) + \chi(r),\qquad \chi(r) = - \frac{1}{v(r)}\frac{d
v(r)}{dr}.  \label{EQ503}
\end{equation}
The equation satisfied by $v(r)$  can be seen to Laguerre equation by
changing variable to $\eta=-\frac{1}{2}\omega r^2$,\begin{equation}
  \eta v{''}(\eta) - \Big(-\eta-(\ell+1) +\frac{1}{2}\Big) v{'}(\eta) -
\frac{R_2}{2\omega} v(\eta ) =0. \label{EQ504}
\end{equation}
Again restricting to polynomial solutions and using $y=-\eta$, we get
\begin{equation}
  v(y) = L_m^{\ell+1/2}(-y)|_{y=\frac{1}{2}\omega r^2}, \qquad \bar{w}(r) =
w_3(r) - \omega r\left[ \frac{\partial_y
L_m^{(\alpha+1)}(-y)}{
L_m^{(\alpha+1)}(-y)}\right]_{y=\frac{1}{2}\omega r^2}. \label{EQ505}
\end{equation}
Substituting $\tilde{w}(r)$ in  \eqref{EQ502} gives the expression for
$\overline{V}^{(+)}(r)$.
We will now find a new superpotential $W_0(r)$ such that
the two extended potentials $\widetilde{V}^{(-)}(r)$ and $\overline{V}^{(+)}(r)$
can be seen to be SUSY partners, viz.,
\begin{eqnarray}
     \widetilde{V}^{(-)}(r) = W_0^2(r) - W_0^\prime(r) \label{EQ510}\\
\text{and    } \overline{V}^{(+)}(r) =  W_0^2(r) + W_0^\prime(r). \label{EQ511}
\end{eqnarray}
%
For this we collect all the equations relating the potentials
$\widetilde{V}^{(\pm)}(r)$ and $\overline{V}^{(\pm)}(r)$ and list them below.
\begin{eqnarray}
    \widetilde{V}^{(+)}(r) &=& V^{(+)}(r) + R_1 \label{EQ508},\\
    \overline{V}^{(-)}(r) &=& V^{(-)}(r) + R_2 \label{EQ509},\\
  \widetilde{V}^{(-)}(r) &=& \widetilde{w}(r) ^2 -  \widetilde{w}(r),
\label{EQ506}
\\
  \overline{V}^{(+)}(r) &=& \overline{w}(r)^2 + \overline{w}^\prime(r).
\label{EQ507}
\end{eqnarray}
Note that (\ref{EQ508}) implies
\begin{equation}
    \phi^2(r,\lambda) +2 w_2(r,\lambda) \phi(r, \lambda) =
       R_1 -  \phi^\prime(r,\lambda).\label{EQ512}
\end{equation}
Similarly, the equation for (\ref{EQ509}) implies that $\chi(r)$ obeys a
relation
\begin{equation}
   \chi^2(r,\lambda) - 2 w_3(r,\lambda) \chi(r, \lambda) =
       R_2 +  \chi^\prime(r,\lambda) \label{EQ513}.
\end{equation}
Using \eqref{EQ510} in the l.h.s of \eqref{EQ506} and substituting $W_0(r) =
w_1(r,\lambda) + \xi(r,\lambda)$ gives
$\widetilde{w}(r) = w_3(r) +\phi(r)$ to get
\begin{eqnarray}
 \lefteqn{ \text{L.H.S. of (\ref{EQ506})}
   =(w_1(r,\lambda)+\xi(r,\lambda))^2 - (w_1^\prime(r,\lambda) + \xi^\prime(r,
\lambda))\label{EQ514}\hspace{1cm}} \\
  &&\hspace*{15mm}= w_1^2(r,\lambda) - w_1^\prime(r,\lambda) + \xi(r,
\lambda)^2 + 2 w_1(r,\lambda)
\xi(r, \lambda) - \xi^\prime(r, \lambda)\label{EQ515}\nonumber\\
  &&\hspace*{15mm}= V(r) -\omega (\ell+3/2) + \xi(r, \lambda)^2 + 2
w_1(r,\lambda)^\prime \xi(r, \lambda) -
\xi^\prime(r, \lambda).\label{EQ516}
\end{eqnarray}
Next, the r.h.s of \eqref{EQ506} is expressed in terms of $\widetilde{w}(r) =
w_2(r) +\phi(r)$ gives
\begin{eqnarray}
 \lefteqn{ \text{R.H.S. of (\ref{EQ506})}
  = (w_2(r,\lambda)+\phi)^2(r,\lambda) - (w_2^\prime(r,\lambda) +
\phi^\prime)\label{EQ517}} \\
  &&\hspace*{15mm}= w_2^2(r,\lambda)- w_2^\prime(r,\lambda) +
\phi^2(r,\lambda) + 2 w_2(r,\lambda)
\phi(r,\lambda) - \phi^\prime(r,\lambda)\label{EQ518}\nonumber \\
  &&\hspace*{15mm}= V(r) +\omega (\ell-1/2) + \phi^2(r,\lambda) + 2
w_2(r,\lambda)
\phi(r,\lambda) - \phi^\prime(r,\lambda).\label{EQ519}
\end{eqnarray}
Equating the expressions in \eqref{EQ516} and \eqref{EQ519} we get
\begin{eqnarray}
   \lefteqn{\xi^2(r,\lambda) + 2 w_1(r,\lambda) \xi(r,\lambda) -
\xi^\prime(r,\lambda)\label{EQ520} \qquad\qquad}\\
   &=& 2\omega(\ell +1)+ \phi^2(r,\lambda) + 2 w_2(r,\lambda) \phi(r,\lambda) -
\phi^\prime(r,\lambda) \label{EQ521} \\
   &=& 2\omega(\ell +1) + R_1-  2\phi^\prime(r,\lambda),   \label{EQ522}
\end{eqnarray}
where, in the last step, (\ref{EQ512}) has been used.
Similarly from \eqref{EQ508} and \eqref{EQ513} we would get
\begin{eqnarray}
   \lefteqn{\xi^2(r,\lambda) + 2 w_1(r,\lambda) \xi(r,\lambda) +
\xi^\prime(r,\lambda)\qquad\qquad} \\
   &=& 2\omega(\ell +1)+ \chi^2 + 2 w_2(r,\lambda) \chi(r,\lambda) -
\chi^\prime(r,\lambda) \label{EQ523} \\
   &=& 2\omega(\ell +1) +R_2 -  2\chi^\prime(r,\lambda).  \label{EQ524}
\end{eqnarray}
Subtracting \eqref{EQ522} and \eqref{EQ524} we get
\begin{eqnarray}
   \xi^\prime (r,\lambda)= \phi^\prime(r,\lambda) -  \chi^\prime(r,\lambda) + A,
\label{EQ525}
\end{eqnarray}
where $A$ is a constant to be fixed. Integrating we get
\begin{equation}
     \xi(r,\lambda) = \phi(r,\lambda) - \chi(r,\lambda) + A r + B.\label{EQ526}
\end{equation}
Both the constants $A$ and $B$ turn out to be zero. This is seen by noting that
the large
$r$ behaviour of the potential  (\ref{vtil_RO}) is correctly reproduced by
$w_1(r,\lambda)$ term in $W_0(r)= w_1(r,\lambda) +\xi(r,\lambda)$. Therefore any
addition of $A r+ B $ with $A, B\ne 0$ in $W_0(r)$ is not possible,
Thus we arrive at the result
\begin{eqnarray}
  W_0(r, \lambda) &=& w_1(r,\lambda) + \phi(r,\lambda) -\chi(r,\lambda).
\label{EQ527}  \\
         &=& \frac{1}{2}\omega r  - \frac{(\ell+1)}{r}
         -\omega r\left[\frac{
\partial_y L_m^{\alpha}(-y)}{
L_m^{\alpha}(-y)}\right]_{y=\frac{1}{2}\omega r^2}
         + \omega r\left[\frac{\partial_y L_m^{(\alpha+1)}(-y)}{
               L_m^{(\alpha+1)}(-y)}\right]_{y=\frac{1}{2}\omega r^2}.
\label{EQ528}
\end{eqnarray}
In this form the superpotential for the extended radial oscillator agrees with
that known in literature \cite{quesne}, \cite{sas_crum}.

The expression for $W_0(r,\lambda)$ can also be obtained by an alternate
approach by noting that  $W_0(r,\lambda)$ is just the superpotential constructed
out of the
ground state wave function of the potential $\widetilde{V}^{(-)}(r)$. The wave
functions for  $\widetilde{V}^{(-)}(r)$ are given by (\ref{EQ341}). Taking
logarithmic derivative of the ground state wave function gives an expression
for $W_0(r,\lambda)$ that matches with (\ref{EQ528}).
The details of this computation  are given in the appendix.

The process of construction of the rational extension of
the radial oscillator potential described in detail above leads to extended
Hamiltonian related to the L1 exceptional orthogonal Laguerre polynomials.
The above  process  of rational extension of potentials by isospectral shift
deformation is summarized in the the left half of the flowchart given in figure
1. In addition,
we also give the steps to construct its shape invariant partner in the right
half.
\subsection{$L2$ and $L3$  rational extensions}
In this section we will briefly summarize our results on rationally extended
potentials obtained when a different route to the rational extension is taken.

\subsubsection*{$L2$ Category}

First we would like to mention that the entire process leading to the
exceptional Laguerre polynomials of $L_1$ category can be repeated by exchanging
the
roles of $w_2(r)$ and $w_3(r)$.  The two  extended potentials, obtained by
these procedures, correspond to the SUSY partner Hamiltonians listed as being
related to exceptional
Laguerre polynomials  of  $L_2$ category by Odake and Sasaki \cite{sas_crum}.

Here, we give the final expressions for the $L2$ exceptional Laguerre
polynomials
\begin{equation}
  \tilde{P}_{n,m}= \frac{1}{(\ell-m+1/2)}\left[ ( \ell-m+1/2) L^{-\ell-3/2}_m(y)
           L^{\ell+1/2}_n(y) + y L^{-\ell-1/2}_m(y)\partial_yL^{\ell+1/2}_n(y)
           )\right]_{y=\frac{1}{2}\omega r^2},
\end{equation}
The corresponding weight function is given by
\begin{equation}
    \mathcal{W}_m =\frac{ r^{\ell}\exp(-\omega r^2/4)}{
L_m^{-\ell-1/2}(y)|_{\frac{1}{2}\omega
r^2}}.
\end{equation}
\setlength{\unitlength}{1mm}

\begin{picture}(190,240)(0,0)
\put(-16,230){\oval(40,20)}
\put(04,230){\line(1,0){20}}
\put(24,210){\line(0,1){40}}

\put(24,250){\vector(1,0){25}}
\put(24,210){\vector(1,0){25}}

\put(-30,233){Superpotentials}
\put(-30,227){$w_k(r),k=1,2,3,4$}
\put(50,245){\framebox(45,10){Exact SUSY $w_1(r), w_4(r)$}}
\put(50,205){\framebox(45,10){Broken SUSY $w_2(r), w_3(r)$}}

\put(70,204){\vector(-3,-1){50}}
\put(70,204){\vector(3,-1){50}}
\put(20,160){\line(0,1){17}}
\put(120,160){\line(0,1){17}}

\put(20,135){\line(0,1){15}}
\put(120,135){\line(0,1){15}}

\put(20,110){\line(0,1){15}}
\put(120,110){\line(0,1){15}}

\put(00, 150){\framebox(40,10){$\widetilde{V}^{(+)}(r)= V^{(+)}(r)+
R_1$}}

\put(00,177){\framebox(40,10){$\widetilde{w}_2(r)=w_2(r)+\phi(r)$}}
\put(00,125){\framebox(40,10){ $\phi=\partial_r\log
L_m^{\alpha}(-y)$}}
\put(00,100){\framebox(40,10){$\widetilde{V}^{(-)}(r)= \tilde{w}^2_2(r) -
\tilde{w}^\prime_2(r)$}}
\put(100, 150){\framebox(40,10){$\overline{V}^{(-)}(r)=V^{(-)}(r)+ R_2$}}
\put(100,177){\framebox(40,10){$\overline{w}_3(r)=w_3(r)+\chi(r)$}}
\put(100,125){\framebox(40,10){ $\chi=\partial_r\log
L_m^{\alpha+1}(-y) $}}
\put(100,100){\framebox(40,10){$\overline{V}^{(+)}= \overline{w}^2_3(r)  +
\overline{w}^\prime_3(r)$}}
\put(30, 68){ \framebox(75,15){ $\widetilde{V}^{(-)}(r)$ and
$\overline{V}^{(+)}(r)$
are SUSY partners}}
\put(60,130){\fbox{$~y=\frac{1}{2}\omega r^2~$}}
\put(20, 100){\vector(3,-1){50}}
\put(120,100){\vector(-3,-1){50}}

\put(35,198){$w_2(r)$}
\put(95,198){$w_3(r)$}
\put(65,60){ Fig.1. }
\end{picture}

\subsection*{$L3$ Category}
The method of spectral shift deformation when applied to the superpotentials
$w_2(r)$ and $w_3(r)$
yielded rational extensions of the radial oscillator potential. These potentials
were shown to have solutions in terms of the $L1$ and $L2$ $X_m$ exceptional
Laguerre polynomials. The above process can be repeated with the superpotentials
$w_1(r)$ and $w_4(r)$ and we can construct further rational extensions of the
radial oscillator.  From the table it is clear that $w_4(r)=-w_1(r,
(l+1)\rightarrow l)$. It is seen that the isospectral shift deformation, using
these two superpotentials leads to two new rational extensions of which one is a
singular potential and not interesting in the present context. We look at the
other potential which is non-singular for certain values of the corresponding
parameters. In general, it has singularities both in the complex plane and the
physical region. These singularities are governed by the zeros of the polynomial
in the corresponding $\phi(r)$. The explicit form of the new generalized
rational extension and its polynomial solutions are presented here for the first
time.

The rational extension obtained by performing the isospectral shift deformation
using $w_1(r)$
and demanding $\widetilde{V}^{(+)}_1(r)=V^{(+)}_1(r)+R_3$ gives
\begin{eqnarray}\lefteqn{
\widetilde{V}^{(-)}_m(r)=\left[V^{(-)}(r)+2\omega
r\left(\frac{\omega r}{2}-\frac{\ell+1}{2}\right)
\frac{\partial_yL^{\alpha}_m(-y)}{
L^ { \alpha}_m(-y)}\right.
}\hspace{12cm}&&\nonumber\\ \hspace{3cm}- \omega r
\partial_y\left(\omega
r\frac{\partial_yL^{\alpha}_m(-y)}{L^{\alpha}_m(-y)}\right)+
2\left.\left(\omega
r\frac{\partial_yL^{\alpha}_m(-y)}{L^{\alpha}_m(-y)}\right)^2\right]_{y=\frac{1
} {2}\omega r^2} &&,
\end{eqnarray}
where $\alpha=-l-3/2$ and again $m=1,2,3...$. The corresponding eigenfunctions
are given by
\begin{equation}
\tilde{\psi}^-_{n,m}(r)= \left(\frac{y^{(l+2)/2} \exp(-\frac{1}{4}\omega
r^2)}{L^{\alpha}_m(\frac{1}{2}\omega r^2)}\right)\tilde{P}_{n,m}(r),
\end{equation} 
where
\begin{equation}
\widetilde{P}_{n,m}(r)=\left[\omega r
L^{-\alpha+1}_n(y)L^{\alpha}_m(-y)+\frac{2}{r}(m-l-\frac{3}{2})L^{\alpha-1}
_m(-y)L^{\alpha}_n(y)\right]_{y=\frac{1}{2}\omega r^2},
\end{equation} 
are polynomials orthogonal with respect to the weight function 
\begin{equation}
\mathcal{W}_m(r)= \frac{r^{\ell+2} \exp(-\omega
r^2/4)}{L^{\alpha}_m(-y)|_{y=\frac{1}{2}\omega r^2}}
\end{equation} 
in the interval $0< r< \infty$.
The weight function has singularities governed by the zeros of the polynomial
$L^{\alpha}_m(-\omega r^2/2)$. From the Kienast-Lawton-Hahn's theorem on the
zeros of the Laguerre polynomials \cite{sze}, \cite{erd}, the polynomial
$L^{\alpha}_m(-\omega r^2/2)$ will not have any zeros on the positive real line
as the
argument of the polynomial is strictly negative. But the polynomial will have
one negative zero if $m$ is odd and $\alpha$ lies in the range
$(-m-\frac{1}{2})<\alpha<-m$.
Thus for even $m$ and $\alpha$ not lying in the above given range, the weight 
function is non-singular in the physical range. Thus for all such values of $m$ 
and $\alpha$, $\tilde{V}^{(-)}(r)$ has polynomial solutions. From the above
discussion it is
clear that the polynomials $\tilde{P}_{n,m}(r)$ are orthogonal and form a 
complete set. In addition, the polynomials start with a degree $m>0$. Thus we 
have a new set of exceptional polynomials and following the nomenclature in the 
literature, we name them the $L3$ exceptional polynomials. The discussion of
the
properties of this new series of the polynomials falls outside the domain of 
this paper and will be taken up else where. 

Thus the isospectral shift deformation using $w_1(r)$ leads to a rational 
extension of the radial oscillator, whose solutions are in terms of new set of 
exceptional polynomials named as the $L3$ series. Rational extensions of radial
oscillator having solutions
other than $L1$ and $L2$ Laguerre EOPs have been discussed in  \cite{sigma} and
\cite{gran2}.

\section{Trigonometric P\"{o}schl-Teller Potential}
The trigonometric  P\"{o}schl-Teller potential is given by
\begin{equation}
  V(x) = A(A+1) \cosec^2x + B(B+1) \sec^2 x, \,\,\, A >-\frac{1}{2}, B >
-\frac{1}{2}
\label{EQ601}.
\end{equation}
The form of the superpotentials is given by
\begin{equation}
  w(x,\sigma)= a \cot x - b \tan x\label{EQ602}
\end{equation}
with $\sigma$ collectively denoting the parameters $\{a,b\}$. The possible
values of the parameters are
\begin{equation}
     a= A, -A-1,\quad b = B, -B-1\label{EQ603}
\end{equation}
and the four solutions of the QHJE correspond to the following choices
\begin{equation}
\begin{array}{ccll}
 \langle1\rangle & E = -(A+B)^2,~~~~  &\qquad& \sigma=\{A,B\} \\
 \langle2\rangle & E = -(1+A-B)^2, &\qquad& \sigma=\{-A-1,B\} \\
 \langle3\rangle & E = -(1-A+B)^2,  &\qquad& \sigma=\{A,-B-1\}     \\
 \langle4\rangle &E = -(2+A+B)^2, &\qquad& \sigma=\{-A-1,-B-1\}\label{EQ80}
\end{array}
\end{equation}
The corresponding solutions will be denoted as $w_k, k=1,2,3,4$, respectively.
With the mapping $\tau$ defined as
$$ \tau\{A, B\} = \{A+1, B+1\},$$
the shape invariance property  is obvious from
$w_4(x,\lambda)= - w_1(x,\tau(\lambda))$ and from $w_3(x,\lambda)= -
w_2(x,\tau(\lambda))$.

\subsection{Rational Extension of Trigonometric P\"{o}schl-Telller Potential}

We note that the first solution in \eqref{EQ80} corresponds to the exact SUSY
case and to a
normalizable ground state solution. Solutions $\langle2\rangle$ and
$\langle3\rangle$ in \eqref{EQ80} correspond to
broken SUSY and will be used to construct rational
extensions with solutions correspondng to $J1$ and $J2$ exceptional Jacobi
polynomials \cite{quesne}, \cite{sigma}, \cite{sas_crum}.

To keep our intermediate steps general we take
\begin{equation}
w(x,\sigma)= a\cot x - b \tan x
\end{equation}
and will substitute suitable values for the constants $\sigma=\{a,b\}$ at the
end. The first extension process is as follows.
\begin{eqnarray}
\widetilde{V}^{(\pm)}
&=& \widetilde{w}(x)^2 \pm \widetilde{w}^\prime(x)\\
\widetilde{w}(x) &=& w(x,\sigma) + \phi(x),
\end{eqnarray}
where $\widetilde{w}(x)$ is determined by demanding
\begin{equation}
\widetilde{V}^{(+)}(x) = V^{(+)}(x)+ K,
\end{equation}
where $K$ is a constant. $\phi(x)$ satisfies the Riccati equation
\begin{equation}
  \phi^2(x) + 2 w(x) \phi(x) + \phi^\prime(x)=K .
\end{equation}
This equation can be linearised by writing
\begin{equation}
\phi(x) = \frac{u^\prime(x)}{u(x)}
\end{equation}
and the equation for $u(x)$ turns out to be
\begin{eqnarray}
  u^{\prime\prime} + 2 w(x,\sigma) u^\prime - K u = 0\\
  u^{\prime\prime} + 2 (a \cot x - b \tan x )u^\prime - K u = 0.
\end{eqnarray}
A point canonical transformation  $y=\cos 2x$ leads to
the equation
\begin{equation}
  (1-y^2) \frac{d^2u}{dy} +[b-a + (a+b+1)y ] \frac{du}{dy} - K_1 u =0
\end{equation}
where $K_1=K/4$ is a constant. This equation has polynomial solutions
with $K_1=N(N+\nu+\mu+1)$ with $\{\nu,\mu\}=\{a-1/2, b-1/2\}$. With this choice
of
$K_1$ the polynomial solution coincides with the Jacobi  polynomial
$P_N^{(\nu,\mu)}(y)$. Therefore, we get
\begin{eqnarray}
  \phi(x,\lambda)
  &=& \frac{1}{u}\Big(\frac{dy}{dx}\Big) \Big(\frac{du(y,\lambda)}{dy}\Big)\\
  &=& -(\sin 2x) \Big[\frac{d}{dy}\log
P_N^{(a-\frac{1}{2},b-\frac{1}{2})}(y)\Big]_{y=\cos2x}.
\end{eqnarray}

One can obtain other rationally extended potentials $\bar{V}^{(+)}(x)$ by means
of a second extension process, as done in the case of radial oscillator. This is
done by writing
\begin{equation}
  \bar{w}(x) = w(x,\sigma) + \chi(x)
\end{equation}
and fixing $\chi(x)$ by demanding $\bar{V}^{(-)}(x)= V^{(-)}(x) + \bar{K}$,
which leads to
the following equations for $\chi(x)$
\begin{equation}
   \chi(x)= - \frac{\bar{u}^\prime(x)}{\bar{u(x)}}\\
  \bar{u}^{\prime\prime} - 2 (a \cot x - b \tan x )\bar{u}^\prime - \bar{K}
\bar{u} = 0.
\end{equation}
Here again $\sigma=\{a,b\}$ and can take one of the four possible values listed
in
\eqref{EQ80}.

As in the case of radial oscillator, the two superpotentials $w_2(x)$, $w_3(x)$
correspond to broken SUSY and will give rise
to two Shape invarinat nonsingular extended potentials and their supersymmetric
partners. We obtain
two more rational extensions and their partners using the other two
superpotentials $w_1(x)$, $w_4(x)$ and
some of these are likely to be singular. Thus starting with the
four superpotentials $w_k(x),k=1,2,3,4$, in all, there are eight possible
extensions obtained by the two processes outlined above and their SI is
guaranteed by
the ansatz that led to equations for the isospectral shift
deformation.

The rational extensions and their partners, obtained from
$w_2(x)$ and $w_3(x)$ will correspond to the  Hamiltonians related to  $J1$, and
$J2$ exceptional
Jacobi polynomials \cite{quesne}, \cite{sas_crum}. Of the remaining  four
extensions, some
will be singular potentials and some others will be regular potentials. 
A detailed study is  needed to make a definite conclusive statement
whether other regular extended potentials are related to the exceptional
Jacobi polynomials already found \cite{gran_dbt} or as in the case of radial
oscillator, will
lead to new  exceptional Jacobi polynomials. Further investigation in this
direction, along with some other related studies is in progress and will be
reported elsewhere.


***
\section{Concluding Remarks}
In this paper we have discussed a procedure to systematically construct
rational extensions of SIPs. The extended potentials are designed 
to be shape invariant. {\bf Our treatment makes it possible to construct an
extension which interpolates
between various potentials related to $X_m-$ exceptional polynomials for 
different values of $m$. So for example, a solution of \eqref{EQ37} with
arbitrary value of the
constant $R$ will give rise to such an extended potential, which for special
values of $R$  will reduce to potentials related to $X_m$ exceptional
polynomials for various $m$ values.}

The method can be used to construct rational extenions of all known SIPs
\cite{khare_book}. Rational extensions of the
radial oscillator and the trigonometric DPT have been constructed explicitly. In
the case
of radial oscillator, we obtained the well studied rational extensions with $L1$
and $L2$
exceptional Laguerre polynomials as solutions. In addition, we have presented
another set of infinite
generalized rational potentials whose solutions involve a new set of exceptional
Laguerre polynomials,
whose generalized expressions are given for the first time.

In the case of the DPT potential, we obtain the extended superpotentials from
which rational extensions of
DPT can be constructed. It is clear that this method leads to new rational
extensions other than those
discussed in literature, hence further study is required to see if these lead to
further new types of
exceptional Jacobi polynomials. It would be interesting to investigate further
generalizations of our method and see
where they lead to.

{\bf Acknowledgments} S S R acknowledges useful discussions with Srinivas Rau
and
thanks the Department of Science and Technology (DST)/SERB, India,
for financial support under the fast track scheme for young scientists (D. O. 
No: SR/FTP/PS-13/2009). Most of this work was done in the above mentioned 
project executed in the Hyderabad University. This work become possible as
large number of previous research was available on the archives
http://xxx.arxiv.org/. The
authors thank all the researchers who make their  manuscripts available on the
archives and the people who maintain the archives.

\section*{APPENDIX}
In this appendix we will give details of \eqref{psithil} and an alternate
derivation of \eqref{EQ528}.

The superpotential and the the Hamiltonian for the radial oscillator are given
by
\begin{eqnarray}
    w(r) &=& \frac{1}{2}\omega r - \frac{\ell+1}{r} \label{A01}\\
    H^{(-)}&=& p^2 + \frac{1}{4}\omega^2r^2 + \frac{\ell(\ell+1)}{r^2}-
(\ell+\frac{3}{2})\omega  \label{A02}.
\end{eqnarray}
The energy eigenvalues and the eigenfunctions are
\begin{equation}
    E_n = 2n \omega; \quad \psi_n^{(-)}(y) = N y^{(\ell+1)/2} \exp(-y/2) L
_n^{\ell+1/2}(y),\label{A03}
\end{equation}
where $y=\frac{1}{2}\omega  r^2$.

The eigenfunctions for $H^{(+)}(r)$, as obtained  from the above expressions, by
replacing  $\ell\to \ell-1$ are
\begin{equation}
  \psi_n^{(+)}(y)= N y^{\ell/2} \exp(-y/2) L _n^\alpha(y), \qquad \alpha=(\ell
-1/2). \label{A04}
\end{equation}
A rational extension was obtained by starting from $w_2= \frac{1}{2}\omega r+
\ell/r$ and introducing $\widetilde{w}(r) = w_2(r) + \phi(r)$. We demanded that
$\phi(r)$ satisfy  $\widetilde{V}^{(+)}(r)= V^{(+)}(r)+R_1$ and be a rational
function
of $r$. This led to
\begin{equation}
  \widetilde{w}_{2,m}(r) = w_2(r) + \omega r\left(\frac{\partial_y
L_m^{\alpha}(-y)}{ L_m^\alpha(-y)}\right).
\end{equation}
The eigenfunctions of the extended potential $\widetilde{V}^{(-)}(r)$
can be computed using the intertwining relation
\begin{equation}
  \widetilde{\psi}_{n,m}^{(-)}(r) = \Big(- \frac{d}{dr} +
                       {\widetilde{w}}_{2,m}(r)\Big)\widetilde{\psi}^{(+)}_n(r).
                       \label{EQ105}
\end{equation}
For this purpose first note that
$\widetilde{\psi}^{(+)}_n(r)={\psi}^{(+)}_n(r)$ of \eqref{A04}, which leads to
\begin{eqnarray}
  \frac{d}{dr} \widetilde{\psi}^{(+)}_{n,m}(r)
  &=& \psi^{(+)}_n (r)\Big(\frac{d}{dr}\log\psi^{(+)}_n(r)   \Big)
\\
  &=& {\psi}^{(+)}_n(y) \Big(\frac{dy}{dr}\Big)
\Big(\frac{d}{dy}\log\psi^{(+)}_n (y)      \Big) \\
  &=& {\psi}^{(+)}_n(y) \times \omega r\Big[ \frac{\ell}{2y}
-\frac{1}{2} + \frac{\partial_y L_n^{\alpha}(y)}{L_n^{\alpha}(y)}
\Big].
  \end{eqnarray}
Use of the above result in \eqref{EQ105}  gives the following
expression for $\psi^{(-)}_{n,m}(r)$
\begin{eqnarray}
  \lefteqn{\psi^{(-)}_{n,m}(r)  =
  \psi^{+}_n(r) (\omega r) \Big[ -\frac{\ell}{2y}
      + \frac{1}{2} -
      \frac{\partial_y L_n^\alpha(y)}{L_n^\alpha(y)} }\\
    && \qquad\qquad +\frac{1}{2} + \frac{\ell}{2y} + \frac{\partial_y
L_m^{(\alpha)}(-y)}{L_m^\alpha(-y)}\Big]\\
&\hspace{6mm}&=\psi^{+}_n(r) (\omega r)\left[
\frac{L_m^{\alpha}(-y)}{L_m^\alpha(-y)}-
\frac{\partial_y L_n^\alpha(y)}{L_n^\alpha(y)}  +1 \right].\label{EQ111}
\end{eqnarray}
Using the identity
 \begin{equation}
   \partial_x L_m^{\alpha}(x) = L_m^\alpha(x) - L_m^{(\alpha+1)}(x),
 \end{equation}
 we get
\begin{equation}
   \widetilde{\psi}_{n,m}^{(-)}(r)
   = (\omega r)\left(\frac{L_m^{\alpha+1}(-y)}{L_m^\alpha(-y)} -
         \frac{\partial_y
L_n^{\alpha}(y)}{L_n^\alpha(y)}\right)_{y=\frac{1}{2}\omega r^2}
          \widetilde{\psi}_n(r).     \label{eq114}
   \end{equation}
The final expression for the wave function $\tilde{\psi}^{(-)}_{n,m}(r)$, as
obtained by using the equations \eqref{EQ111} and \eqref{eq114}
is
\begin{equation}
  \tilde{\psi}^{(-)}_{n,m}(r) = r^{\ell/2}
\exp(-\frac{1}{4}\omega r^2)\left[
\frac{L_m^{\alpha+1}(-y)L_n^\alpha(y)-\partial_y
L_n^{\alpha}(y)L_m^\alpha(-y)}{L_m^\alpha(-y)}\right]_{y=\frac{1}{2}\omega r^2}
\end{equation}
where $\alpha=\ell-1/2$. This is the result in \eqref{psithil} and
the expression for the eigenfunctions can be used to have an alternate
derivation of \eqref{EQ528}.\\

Setting $n=0$ in \eqref{eq114} gives
\begin{equation}
  \widetilde{\psi}_{0,m}^{(-)}(r) = (\omega
r)\left(\frac{L_m^{\alpha+1}(-y)}{L_m^\alpha(-y)}
\right)_{y=\frac{1}{2}\omega r^2} \widetilde{\psi}_{0,m}^{+}(r). \label{w0}
\end{equation}
The desired superpotential can be obtained as
\begin{equation}
  W_0(r) = -\frac{d}{dr}\Big(\log \widetilde{\psi}_{0,m}^{(-)}(r)
\Big).
\end{equation}
Substituting the expression in \eqref{w0} in the above equation gives
\begin{eqnarray}
  W_0(r)
  &=& - \frac{d}{dr}\Big( \log
\widetilde{\psi}^{(-)}_{n,m}(r)\Big)\Big|_{n=0}\\
  &=& -\frac{1}{r}-\frac{d}{dr} \log \widetilde{\psi}^{(+)}_0(r)+ \omega r\left[\frac{\partial_y L_m^\alpha(-y)}{L_m^\alpha(-y)}-\frac{\partial_y L_m^{(\alpha+1)}(-y)}{L_m^{(\alpha+1)}(-y)} \right]_{y=\frac{1}{2}\omega r^2}. \\
&=&\frac{1}{2}\omega r -\frac{\ell+1}{r} + \omega r\partial_y \log \Big(\frac{
L_m^\alpha(-y)}{ L_m^{\alpha+1}(-y)}\Big)_{y=\frac{1}{2}\omega r^2}-.
\end{eqnarray}
which agrees with \eqref{EQ528}.



\begin{thebibliography}{sasaki9sigma}

\bibitem{kam1} D. G\'{o}mez-Ullate D, N. Kamran  and R. Milson,  J. Math. Anal.
Appl. 359 (2009) 352-367;
(arXiv:0807.3939).

\bibitem{kam2} D. Gomez-Ullate, N. Kamran and R. Milson, J Approx Theory 162
(2010) 987-1006; (arXiv:0805.3376).

\bibitem{quesne} C. Quesne, J. Phys. A: Math. Theor. 41 (2008) 392001-392006.

\bibitem{sasaki} S. Odake and R. Sasaki, Phys. Lett. B 679 ( 2009) 414-417; (arXiv:0906.0142).

\bibitem{khare_book} F. Cooper, A. Khare and U. P. Sukhatme, Supersymmetric
quantum mechanics, World Scientific Publishing Co. Ltd. Singapore, 2001.

\bibitem{kam3} D. Gomez-Ullate, N. Kamran and R. Milson, J. Phys. A 37 (2004)
1780-1804; (arXiv:quant-ph/0308062).

\bibitem{kam4} D. Gomez-Ullate, N. Kamran and R. Milson, J. Math. Anal. Appl.
387 (2012) 410-418;(arXiv:1103.5724).

\bibitem{sigma} C. Quesne, Sigma 5 (2009) 084;(arXiv:0807.4650).

\bibitem{sasaki1} S. Odake and R. Sasaki, Phys. Lett. B 682 (2009) 130-136; (arXiv:0909.3668)

\bibitem{sas_crum} R. Sasaki , S. Tsujimoto and A. Zhedanov, J. Phys. A: Math.
Theor. 43 (2010) 315204; (arXiv:1004.4711).

\bibitem{kam_dar} D. Gomez-Ullate, N. Kamran, R. Milson , J. Phys. A 43 (2010)
434016.

\bibitem{gran1} Y. Grandati, Physics Letters A 376 (2012) 2866-2872.

\bibitem{gran2} Y. Grandati,  Ann. Phys. 326 (2011) 2074-2090.

\bibitem{gran_dbt} B. Bagchi, Y. Grandati and C. Quesne, arXiv:1411.7857.

\bibitem{ho} C.-L. Ho, J. Phys. A: Math. Theor. 52 (2011) 122107-122108.

\bibitem{dutta} D. Dutta and P. Roy, J. Phys. A: Math. Theor. 51  (2010) 042101.

\bibitem{chait} Chaitanya K V S, Sree Ranjani S, Panigrahi P K, Radhakrishnan R
and Sriinivasan V, accepted for publication in Pramana,  arXiv:1009.1944.

\bibitem{sasaki_sigma} C.-L. Ho , S. Odake and R. Sasaki, SIGMA 7 (2011) 107.

\bibitem{gomez1} D. G\'{o}mez-Ullate, F. Marcell\'{a}n, R. Milson, J. Math.
Anal. Appl. 399 (2013)  480 - 495; (arXiv:1204.2282).

\bibitem{gomez2} D. Gomez-Ullate, N. Kamran, R. Milson,  Found. Comput. Math. 13 (2013) 615-656; (arXiv:1203.6857). 

\bibitem{ces} G. Junker and P. Roy P Ann. Phys. 270 (1998) 155-177;(quant-ph/9803024).

\bibitem{ged} Gendenshtein L 1983 JETP Lett. 38 356-359.

\bibitem{wit} E. Witten, Nucl. Phys. B188 (1981) 513;

\bibitem{coop} F. Cooper and B. Freedman, Ann. Phys. 146 (1983) 262-288.

\bibitem{lea} R. A. Leacock  and M. J. Padgett,  Phys. Rev. Lett. 50 (1983) 3; R. A. Leacock and  M. J. Padgett, Phys. Rev. D 28 (1983) 2491.

\bibitem{bhalla} R. S. Bhalla, A. K. Kapoor and P. K. Panigrahi,  Am. J. Phys. 65 (1997) 1187-1194; (arXiv: quant-ph/9512018).

\bibitem{es} S. Sree Ranjani, K. G. Geojo, A. K. Kapoor and P. K. Panigrahi,  Mod. Phys. Lett. A 19 (2004) 1457-1468; (arXiv: quant-ph/0211168).

\bibitem{the} S. Sree Ranjani, Quantum Hamilton - Jacobi solution
  for spectra of several one dimensional potentials with special
  properties, thesis submitted to the University of Hyderabad, 2005;
(arXiv:0408036).

\bibitem{sree_eop} S. Sree Ranjani, P. K. Panigrahi, A. Khare, A. K. Kapoor, A. Gangopadhyaya, J. Phys. A: Math. Theor. 45 (2012) 055210.

\bibitem{qes} K. G. Geojo, S. Sree Ranjani and A. K. Kapoor, J. Phys. A: Math. Gen. 36 (2003) 4591–4598; (arXiv: quant-ph/0207036).


\bibitem{sip} R. Sandhya, S. Sree Ranjani and A. K. Kapoor, arXiv:1412.4244.

\bibitem{khare_iso} Khare A and Sukhatme U 1989 J. Phys. A: Math. Hen. 22 2847.

\bibitem{sukatme_iso} Pappademos J, Sukhatme U and Pagnamenta A 1993 Phys. Rev. A 48 3525.


\bibitem{den} P. Dennery and A. kriswicki, Mathematics for Physicists, Dover Publications, New York, 1996.
	
\bibitem{ince} E. L. Ince, Ordinary Differential Equations, Dover Publications Inc., New York, 1956.

\bibitem{sze} G. Szeg\"o, Orthogonal polynomials, American Mathematical Society,
Providence, 1975.

\bibitem{erd} A. Erd\'elyi, W. Magnus, F. Oberhettinger and F. G. Tricomi,
Higher Transcendental functions, McGraw-Hill, New York, 1953.



\end{thebibliography}
\end{document}